\def\fdeg{\hbox{$.\!\!^{\circ}$}}

\documentstyle[12pt,aasms4]{article}

\begin{document}

\title{Atomic Hydrogen and Star Formation \\in the
Bridge/Ring Interacting Galaxy Pair\\
NGC 7714/7715 (Arp 284)}
\author{Beverly J. Smith}
\affil{IPAC/Caltech, MS 100-22, Pasadena CA  91125}
\author{Curtis Struck}
\affil{Department of Physics and 
Astronomy, Iowa State University,
Ames IA  50012}
\centerline{and}
\author{Richard W. Pogge}
\affil{Department of Astronomy,
Ohio State University, 174 W. 18th Ave, Columbus OH 43210}
\vskip 0.8in
To be published in the July 10, 1997 issue of the Astrophysical Journal

\begin{abstract}

We present high spatial resolution 21 cm HI maps of the interacting
galaxy pair NGC 7714/7715.  We detect a 
massive ($2\times10^9$M$_{\sun}$) 
HI bridge connecting the galaxies that is parallel to but offset from the
stellar bridge.  A chain of HII regions traces the gaseous bridge,
with H$\alpha$ peaks near but not on the HI maxima.  
An HI tidal tail is also
detected to the east of
the smaller galaxy NGC 7715,
similarly offset from a stellar tail.
The strong partial stellar ring on the east
side of NGC 7714 has no HI counterpart, but on the opposite side of
NGC 7714 there
is a 10$^9$M$_{\sun}$
HI loop $\sim11$~kpc in radius.
Within the NGC 7714 disk, 
clumpy HI gas is observed associated with star formation regions.
Redshifted HI absorption is detected towards the strong
starburst nucleus.  We compare the observed morphology and gas
kinematics with gas dynamical models in which a low-mass companion has an off-center
prograde collision with the outer disk of a larger galaxy.
These simulations
suggest that the bridge in NGC 7714/7715 is a hybrid between the tidal
bridges seen in systems like M51 and the purely gaseous `splash'
bridges found in ring galaxies like the Cartwheel.  The offset between
the stars and gas in the bridge may be due to dissipative cloud-cloud
collisions occuring during the impact of the two gaseous disks.

\end{abstract}

\keywords{galaxies: individual (NGC 7714/7715) -
galaxies: interactions - galaxies: kinematics and dynamics -
radio lines: atomic}

\section{Introduction}

It is well-established that 
gravitational encounters between galaxies 
can create tidal tails and bridges 
\markcite{tt72}(Toomre $\&$ Toomre
1972; \markcite{w72}Wright 1972). It is
also well-known that 
head-on collisions 
between unequal mass galaxies
can create rings \markcite{lt76}(Lynds $\&$ Toomre 1976;
\markcite{ts77}Theys $\&$ Spiegel 1977).
Luminous HII regions have been observed in 
extended tidal features 
\markcite{s78}(Schweizer 1978; \markcite{m91}Mirabel et al. 1991, 
\markcite{m92}1992;
\markcite{hv96}Hibbard $\&$ van Gorkom 1996) as well as in
collisional rings \markcite{fh77}(Fosbury $\&$ Hawarden 1977;
\markcite{h95}Higdon 1995;
\markcite{ma95}Marston $\&$ Appleton 1995), indicating that
interactions can trigger 
star formation 
in these peculiar structures.

A particularly interesting class of interacting galaxies
are those pairs containing both a bridge and a collisional ring.
The location of the bridge in a bridge/ring system
may indicate which of
multiple companions collided with the 
ring galaxy, as in the 
Cartwheel
galaxy 
\markcite{h96}(Higdon 1996) and in 
VII Zw 466 \markcite{acs1996}(Appleton
et al. 1996), while
the composition of the bridge may provide clues
to how it formed.
Two different mechanisms are capable of creating bridges
between galaxies.  As discussed by 
\markcite{tt72}Toomre $\&$ Toomre (1972),
tidal forces and spin-orbit coupling during an encounter are
capable of drawing both gas and stars out into a bridge.
A well-known example of a tidal bridge is found
in the M51 system \markcite{tt72}(Toomre $\&$ Toomre 1972).
Tidal bridges have been 
modeled in detail by \markcite{bh91}Barnes $\&$ Hernquist (1991)
and \markcite{mh96}Mihos $\&$ Hernquist (1996) using
a hybrid N-body/hydrodynamical code.
In these simulations, the
impact parameters used are greater than the
disk radii, so an immediate collision between the two
galaxies does not occur.
The second type of bridge, a `splash' bridge, is 
the result of a purely hydrodynamical effect and occurs
during smaller impact parameter encounters
\markcite{acs96}(Appleton
et al. 1996; 
\markcite{s96a}Struck
1996a, \markcite{s96b}1996b).  
In this version of the standard ring galaxy
model, both galaxies 
initially contain gaseous disks.
During the passage of the smaller
galaxy through the larger disk,
gas clouds from the two galaxies
collide, 
creating a bridge between the two galaxies.
In this class of bridge formation models, a stellar counterpart
to the gaseous bridge is not formed
and the expected gas surface density in
the bridge is 
relatively low
\markcite{s96a}(Struck 1996a).
The HI bridges seen in the 
Cartwheel 
\markcite{h96}(Higdon
1996) and VII Zw 466 systems
\markcite{acs96}(Appleton et al.
1996) may be this type of bridge, rather than 
tidal bridges \markcite{s96a}(Struck 1996a; 
\markcite{s96c}Struck
et al. 1996), because
the observed bridges are purely gaseous.

In this paper, we present new 
21 cm HI and narrowband optical images of the bridge/ring
system NGC 7714/7715 (Arp 284), and discuss
them in the context of ring and bridge formation models.
This system differs dramatically from the 
Cartwheel and VII Zw 466 systems in having a prominent
stellar bridge as well as only
a partial ring \markcite{a66}(Arp 1966). 
In addition, NGC 7714
differs from the other two galaxies
by having tidal tails \markcite{a66}(Arp 1966)
and a strong bar
\markcite{bw90}(Bushouse
$\&$ Werner 1990).
This complex morphology may have been caused by
a more off-center encounter than that which 
produced
the more symmetric rings in
the Cartwheel and VII Zw 466.
In \markcite{paperI}Smith $\&$ Wallin (1992; hereafter Paper I),
we used a restricted 3-body model
to simulate the NGC 7714/7715 encounter, and successfully
reproduced the 
observed optical structure with
an off-center (impact parameter r$_{min}$ $\sim$ 0.85 times the
radius of the larger disk) inclined 
collision between two unequal-mass disk galaxies
(M$_2$/M$_1$ $\sim$ 0.3).
NGC 7714/7715 is 
somewhat reminiscent of 
the `Sacred Mushroom' galaxy
AM~1724-622
\markcite{am86}(Arp $\&$ Madore 1986), which
also 
contains a ring and a stellar bridge.
AM~1724-622,
however, does not
have tidal tails and therefore 
may
have undergone a less off-center
(r$_{\min}$ $\sim$ 0.5 r$_{disk}$)
collision than NGC 7714/7715
\markcite{ws94}(Wallin
$\&$ 
Struck-Marcell 1994).

The distribution of star formation in NGC 7714
also differs from that in many ring galaxies.
In contrast to the outer ring in the Cartwheel 
\markcite{fh77}(Fosbury $\&$ Hawarden 1977; 
\markcite{h95}Higdon 1995) and the ring
galaxies surveyed by 
\markcite{ma95}Marston $\&$ Appleton (1995),
the NGC 7714 ring does not have on-going star formation
\markcite{bw90}(Bushouse $\&$ Werner 1990;
\markcite{b93}Bernl\"ohr 1993;
\markcite{g95}Gonz\'alez-Delgado et al. 1995).
The near-infrared colors of the NGC 7714 ring imply
a considerable older stellar component \markcite{bw90}(Bushouse
$\&$ Werner 1990), while the Cartwheel's outer ring is primarily
composed of young stars \markcite{mah92}(Marcum, Appleton, $\&$ Higdon 1992;
Higdon 1995).
The NGC 7714 ring resembles the
inner ring 
of the Cartwheel, which has
an older stellar population \markcite{mah92}(Marcum
et al. 1992) and little star formation 
\markcite{h95}(Higdon 1995), as well
as the ring in AM~1724-622,
which is also composed of mainly older stars
\markcite{ws94}(Wallin
$\&$ 
Struck-Marcell 1994).

Star formation in NGC 7714
is occurring
at the nucleus, near the northern 
end of the bar, and 
at the base of the southwestern tidal tail
instead of in the ring
\markcite{bw90}(Bushouse $\&$ Werner 1990;
\markcite{b93}Bernl\"ohr 1993;
\markcite{g95}Gonz\'alez-Delgado et al. 1995).
The luminous starburst nucleus in this galaxy is 
well-studied
\markcite{f80}(French 1980; \markcite{w81}Weedman et al. 1981;
\markcite{k84}Keel
1984;
\markcite{ds88}de 
Robertis $\&$ Shaw 1988) and is known to contain
Wolf-Rayet stars
\markcite{v85}(van Breugel et al. 1985;
\markcite{g95}Gonz\'alez-Delgado et al. 1995).
Luminous HII regions are also seen in the bridge 
between the two galaxies but not in the companion NGC 7715
\markcite{a66}(Arp 1966; 
\markcite{b93}Bernl\"ohr 1993;
\markcite{g95}Gonz\'alez-Delgado et al. 1995).
NGC 7715 has no optical
emission lines 
\markcite{hms56}(Humanson, Mayall, $\&$ Sandage 1956; 
\markcite{b93}Bernl\"ohr 1993);
spectral
synthesis indicates that it is in a post-starburst stage,
with a large population of A and late-type B stars
\markcite{b93}(Bernl\"ohr
1993).
Most of the far-infrared emission from this system arises
from NGC 7714 \markcite{s93}(Surace et al. 1993).

In our previous study of this system \markcite{PaperI}(Paper I),
we presented
low resolution (40$''$)
21 cm HI maps obtained with the NRAO Very Large Array
(VLA\footnote{The VLA is operated by Associated Universities, Inc.,
under contract with the NSF.})
in the D array configuration.
These observations showed that the atomic hydrogen 
in this system is very
extended, reaching out to a maximum radius of 6$'$ ($\sim$70 kpc).
Between the two galaxies, a gaseous bridge is seen.
In spite of its high star formation rate,
NGC 7714 is relatively weak in the standard
tracer of molecular gas, the millimeter CO (1 $-$ 0) line
(\markcite{sss91}Sanders, Scoville, $\&$ Soifer 1991; 
\markcite{yd91}Young $\&$ Devereux 1991).
The
M$_{H_2}$/M$_{HI}$ ratio in the inner
45$''$ has the unusually low value of 0.75, assuming the standard Galactic
I$_{CO}$/N$_{H_2}$ ratio
\markcite{paperI}(Paper I).
This may be due to the low
metallicity of this system \markcite{w91}(Weedman et al. 1991; 
\markcite{g95}Gonz\'alez-Delgado et al. 1995) rather
than a true deficiency of molecular gas \markcite{PaperI}(Paper I).
Interferometric observations show that
the CO peak in NGC 7714 is a few arcseconds
north of the optical and radio continuum nucleus, and is
extended about 10$''$ east-west
with a spur extending south towards the nucleus
\markcite{i93}(Ishizuki 1993).

The VLA and optical observations of
NGC 7714/7715 are described in Section 2 of this paper, while
the data are described in Section 3.  
To better understand the role of tidal
and hydrodynamical effects
in shaping the gas morphology of this system,
in Section 4 we present a 
hydrodynamical model of a low mass disk
galaxy impacting the outer disk of a larger galaxy.
Implications of the observations and model
are discussed in Section 5, and conclusions given in Section 6.
Table 1 provides some basic information about the two 
galaxies in this system.
A distance to these galaxies of 37 Mpc is assumed
throughout this paper (H$_o$ = 75 km s$^{-1}$ Mpc$^{-1}$).

\section{Observations and Data Reduction}

\subsection{Very Large Array}
 
The new HI data were obtained with
the B and C array configurations of the VLA.
The C Array observations
were made on 1992 May 21, while
the B Array measurements were obtained on
1994 July 10.
The total integration time on NGC 7714/7715 was 8 hours with
the C Array and 7.5 hours with the B Array.
During both observing runs,
3C 286 and 3C 48 were each observed for 20 minutes for flux 
and bandpass calibration,
and 
a nearby 
phase calibrator was observed every 30 minutes for 5 minutes.
On-line Hanning smoothing was used for the observations,
giving 63 channels with a resolution of 10.5 km s$^{-1}$ 
and a total bandwidth
of 655 km s$^{-1}$ centered at 2800 km s$^{-1}$.

Calibration and mapping were accomplished using the NRAO Astronomical
Image Processing System (AIPS).
After calibration, the
first step in 
the data reduction
was to make
a continuum map from 21 line-free channels near the ends
of the bandpass.  This map was then CLEANed \markcite{c80}(Clark 1980) and
the CLEAN components were subtracted from the original UV database.
A set of continuum-free maps were then constructed and
residual continuum emission removed.  The maps were then CLEANed
to remove sidelobes due to the HI emission.
Finally, a correction for primary beam attenuation was made.
To obtain the highest sensitivity possible, 
natural weighting was used for these maps.
A set of maps was obtained from the B Array data alone, giving
a final beamsize of 6\farcs32 $\times$ 5\farcs79
with a position angle of 2\fdeg7
and a noise level of 0.39 mJy beam$^{-1}$ channel$^{-1}$.
A second
set of maps was
made after combining the B and C Array data with the 
D Array data from Paper I.
The final beamsize for the B+C+D Array dataset
is 11\farcs02 $\times$
8\farcs48 with a position angle of $-$36\fdeg6
and noise level of 0.32 mJy beam$^{-1}$
channel$^{-1}$.
The B Array datacube was made with 1024 $\times$ 1024
pixels and 
1$''$ pixel$^{-1}$, while the B+C+D Array maps
have 512 $\times$ 512 pixels with 2$''$ pixel$^{-1}$.
The final B+C+D Array HI 
channel maps are shown in Figure 1.  
Emission is seen in 25 channels, from
2674 km s$^{-1}$ to 2926 km s$^{-1}$.

HI intensity maps were made from these two sets
of channel maps
using the AIPS moment routines.
To determine which pixels from the cube
should be excluded in constructing the final B Array intensity map, 
a scratch copy of the datacube was 
Hanning smoothed to 31.2 km s$^{-1}$ and convolved
with a 7$''$ $\times$ 7$''$ FWHM Gaussian beam.
Only pixels above 0.65 mJy beam$^{-1}$ in this
smoothed cube were included in deriving the final map from the original 
unsmoothed cube.
To search for more extended emission, 
the same method was used on
the higher sensitivity B+C+D Array datacube to construct
a second HI intensity map.
For this second map, the scratch cube was 
Hanning smoothed to 31.2 km s$^{-1}$
and convolved with a 16$''$ $\times$ 16$''$ Gaussian beam.
Data below 0.85 mJy beam$^{-1}$ in the smoothed cube
were excluded in making the moment 0 map.
We also construct a moment~1 HI velocity map
from the 
B+C+D Array dataset.  In deriving this map, we use
a scratch cube
smoothed with 
a 16$''$ $\times$ 16$''$
FWHM Gaussian and a
1.1 mJy beam$^{-1}$ flux cutoff, with no velocity smoothing.

\subsection{Narrowband Optical Imaging}

NGC 7714/7715 was observed 
on the night of 1991 October 8
with the 1.8m
Perkins Telescope of the Ohio State
and Ohio Wesleyan Universities at Lowell
Observatory. 
The Ohio State
University Imaging Fabry-Perot Spectrograph 
was used in its direct imaging mode with
the Lowell NSF TI 800 $\times$ 800 CCD detector.
This setup
gives 0.49$''$ pixel$^{-1}$.
Two 50 $\AA$ interference filters at
6660 $\AA$ and 6560 $\AA$ were used to produce
on- and off-H$\alpha$ images, respectively.
The total exposure time per filter was 1800 sec.
The seeing was 1\farcs7.

The images were flat-field and de-biased
in a standard fashion, and then registered,
scaled, and subtracted to produce a pure 
emission-line image.  
Final registration
was accomplished using stellar positions
measured by the McDonald Observatory
PDS machine, as in \markcite{paperI}Paper I.
Since this image was obtained during
nonphotometric conditions, it was
approximately flux calibrated using the 
total H$\alpha$ flux of the system from
\markcite{g95}Gonz\'alez-Delgado et al. (1995).

\section{Observational Results}

\subsection{The Narrowband Optical Images}

The final narrowband
continuum and H$\alpha$ 
images are shown in Figures 2a
and 2b, respectively.
Figure 2b
is consistent
with previous H$\alpha$ maps of the system 
\markcite{bw90}(Bushouse
$\&$ Werner 1990;
\markcite{b93}Bernl\"ohr 1993;
\markcite{g95}Gonz\'alez-Delgado et al. 1995).
Massive star formation is observed
in the nucleus of NGC 7714;
this nuclear HII region
complex
is asymmetric, with
an eastern extension.
Luminous HII regions lie along an arc
at the northern end of the NGC 7714 bar and at the
base of the southwestern tail, but
no H$\alpha$ emission is detected from
the NGC 7714 ring.
A prominent chain of HII regions is
visible in the bridge connecting the two galaxies.
No HII regions are seen in NGC 7715.
In the optical continuum image (Figure 2a), both galaxies
are seen, and the NGC 7714 ring and bar
are visible. 

\subsection{The Radio Continuum Map}

The B Array continuum map is shown in Figure 3.
The center of NGC 7714 and a triple
source to the northeast of NGC 7715 (a background object) are seen
in this plot; NGC 7715 is not
detected in the radio continuum.  
NGC 7714 has a bright central core surrounded
by diffuse extended emission.
The peak and total
20 cm B Array continuum flux density for NGC 7714 
are 26 mJy beam$^{-1}$ and
77 mJy, compared to 59 mJy beam$^{-1}$
and 71 mJy, respectively, for the D Array data \markcite{paperI}(Paper I).
The HII region complex at the base of the southwestern
tail of NGC 7714 is resolved in this map.
It has a total 20 cm continuum flux of 3.5 mJy.

The positions and peak flux densities 
of the three peaks in the background source are tabulated in Table 2.
Source $\#$2 is coincident with
the z = 1.87 
18th magnitude optical quasar 2333+0154 \markcite{b94}(Bowen
et al. 1994).
The double-lobed radio structure of this object was previously
noted by \markcite{nhg89}Neff, Hutchings, $\&$ Gower (1989).
The total B Array flux density for this background object is
176 mJy, compared to 157 mJy from the D Array \markcite{paperI}(Paper I).

\subsection{The HI Intensity Maps}

In Figure 4a, we present the naturally-weighted B Array
HI intensity map, while in Figure 4b,
the naturally-weighted B+C+D Array HI intensity map
is given.  
Figures 5a and 5b show
an expanded view of 
the higher resolution B Array HI intensity map
superposed on 
the narrow band optical continuum image and the H$\alpha$ image,
respectively.
To search for low level optical continuum emission, we have
smoothed the red Digitized Sky 
Survey\footnote{The Digitized Sky Survey, a compressed digital
form of the Palomar Observatory Sky Atlas,
was produced at the Space Telescope Science
Institute 
under U.S. Government grant NAG W-2166. 
The National Geographic Society - Palomar Observatory Sky Atlas (POSS-I)
was made by the California Institute of Technology with grants from the
National Geographic Society.}
data for this region 
to 12$''$ resolution.
It is overlaid on the B+C+D Array HI intensity map in Figure 6.

The HI morphology of this system
is very disturbed and the gas extends well beyond the main optical
disks of the galaxies.  Particularly prominent in these maps is an HI
bridge between the two galaxies.
This bridge is offset 
$\sim$10$''$ (1.8~kpc) to the north of the optical bridge (Figure 6).
The H$\alpha$ sources 
apparent in the bridge lie near,
but not on,
the HI peaks (Figure 5b).  Massive star formation is occuring in
the gaseous bridge rather than in the more
southern stellar bridge. This effect is also
seen in the Arp Atlas \markcite{a66}(Arp 1966) photograph
of NGC 7714; the H~II regions lie
to the north of the smoother
stellar bridge.  

Tidal HI is seen in other locations as well.
Strong HI is observed in a prominent
countertail
to the east of NGC 7715.  This feature is hereafter 
named the `eastern cloud'.
HI is relatively weak at NGC 7715; in the less
sensitive B Array data a gap is observed between the eastern
cloud and the bridge.
In the higher sensitivity B+C+D Array data this
gap is filled in with more diffuse gas.
As at the bridge,
the 
gas
appears skewed to the north relative to
the optical light
throughout both NGC 7715 and the eastern tail.

On the other side of the system, there is a bright arc or loop
of HI 1$'$ northwest of NGC 7714 (the `northwestern loop').
The smoothed POSS image (Figure 6) shows a possible
optical counterpart to
this HI loop.
Finally, to the far west, there are numerous HI clouds
that extend
several arcminutes from the center of NGC 7714 (Figure 4b).

Our data also provide a closer look at NGC 7714 itself.
Figures 4a and 4b show 
an apparent HI hole at the optical
nucleus of NGC 7714;
this depression may be due to HI absorption (see
Section 3.5).  
In the disk of NGC 7714, the HI is not confined
to a bar or to an 
HI ring
tracing the optical ring
(Figure 5a).
Instead, the gas is quite clumpy,
with two HI holes at the outer edge of the stellar
ring and another depression near the southern 
end of the bar (Figure 5a).
South of the ring there is an arc of gas extending
to the southeast
(the `southeastern arc').
Particularly
large clumps of gas
are seen near the star formation regions northwest
and southwest of
the nucleus (the `northwestern' and `southwestern clumps').
The southwestern clump lies at the base of the
optical tail.
In these regions, the H$\alpha$ peaks are close
to but not coincident with the HI peaks.
There is also an HI peak inside the ring (Figure 5a).
There is a small HII region outside of this
HI clump, on the inner edge of the ring (Figure 5b).
This clump of gas may
be foreground or background bridge
material rather than disk gas, and may 
not lie in the same plane as the ring.
Its presence therefore 
complicates the search for a gaseous
counterpart to the ring.

We find a
total HI flux for this system of 21.8 Jy km s$^{-1}$,
comparable to the 10$'$ aperture single dish measurements of
17.9 Jy km s$^{-1}$ \markcite{ps74}(Peterson $\&$ Shostak 1974)
and 21.0 Jy km s$^{-1}$ \markcite{b87}(Bushouse 1987).
The HI masses and peak column densities
of the various structures
are
given in Table 3. 
More than half of the total atomic gas in the system
lies outside of the optical disks.  The HI masses
of
the bridge and the 
western loop are especially large, greater than 10$^9$ M$_{\sun}$.
The peak HI column density in the bridge,
5.4 $\times$ 10$^{21}$ cm$^{-2}$ in a 6$''$ beam,
is particularly high.

\subsection{The HI Velocity Structure}

In Figure 7, the HI velocity profiles for 
various structures 
in the NGC 7714/7715 system
are shown.  The peak velocities
and FWHM line widths of these
features are included in Table 3.  
The total profile agrees with 
the single dish observations of \markcite{ps74}Peterson
$\&$ Shostak (1974).
The gas in the bridge and eastern cloud is quite confined in
velocity, as is typical of tidal features, while
the western loop has a relatively broad velocity profile.
The profiles for the individual clumps in the
NGC 7714 disk are relatively
narrow.

In Figures 8a and 8b, we plot the
HI velocity field 
with the narrowband optical
continuum map and the B+C+D Array HI intensity map, respectively.
East to west across the system from the eastern cloud
through
the bridge, a shallow velocity gradient of 3.5 km s$^{-1}$ kpc$^{-1}$
is observed, with the gas in the eastern cloud being
more blueshifted.
At 
NGC 7715 the gas is
more redshifted than expected by this trend (Figure 8a).
Across the optical body of NGC 7715,
the velocity field suggests rotation with a line of
nodes parallel to the major axis of the galaxy
and the western side redshifted (Figure 8a).
A velocity change of 50 km s$^{-1}$ is observed
over 20$''$ (3.6 kpc), and the systemic velocity
is 2800 km s$^{-1}$, equal to that of NGC 7714.
Thus the kinematics of the gas in the immediate
vicinity of NGC 7715 appear
distinct from those of the surrounding extended
gas.

In the disk of NGC 7714, the velocity field indicates the presence
of an inclined rotating
disk of gas, with more redshifted
gas to the northwest.
The line of nodes is approximately
parallel to the bar, at a position angle of
$\sim$132$^{\circ}$ in the inner disk, decreasing
slightly 
with increasing radius.
The velocity field shows maxima and minima beyond
the end of the bar at radii of 15$''$ $-$ 20$''$,
suggesting that the rotation curve turns over at 
that radius.  Relative to the systemic velocity,
the maximum rotational velocity uncorrected
for inclination is 60 km s$^{-1}$.

The most redshifted gas in the system is in the
northwestern loop.
East to west along this loop
the velocities increase until the point at
which the loop bends southward.  After that
location the gas becomes increasingly blueshifted
as the loop arcs southward and connects back
to the main galaxy.
The position of maximum redshift does not lie
along the line of nodes defined by the inner disk,
but instead is rotated 20$^{\circ}$ to the southwest.
In the far western HI clouds, the velocity
gradient is such that the most northern gas
is more redshifted than that in the south.

In Figures 8a and 8b, significant deviations from
circular motion are observed.
Note the particularly strong peculiar motions
at the location of the gaseous bridge (Figure 8b).
Figure 8a also shows 
some curvature in the isovelocity
contours at the stellar ring.
In addition, there are deviations near the southwestern
HI clump, where the gas is more redshifted than
expected (Figure 8b).
These deviations may be caused by a combination
of non-planar and radial motions.
It is probable that the gas in the bridge
lies in a different plane than the gas in the disk
of NGC 7714, and the disk
of NGC 7714 itself may be warped.  
Second, radial motions may be present along
the bridge, in the ring, and along the tails.
Because of the uncertain
geometry of the system it is difficult to
distinguish 
between these possibilities.

From the HI velocity field, we 
have derived a rotation curve for the
inner disk of NGC 7714 (Figure 9).
We 
used a tilted ring model
as in \markcite{b89}Begeman (1989) and
weighted the data 
by the cosine of the
azimuthal angle with maximum weighting
along the major axis.
Because of the 
observed deviations from circular motion,
we do not iteratively fit all the required parameters
for this curve.  
Instead, we assume that the dynamical center
is coincident with the radio continuum nucleus 
\markcite{c90}(Condon
et al. 1990; Table 1), the systemic velocity
is 2800 km s$^{-1}$, and the line of nodes
is fixed at a position angle of 132$^{\circ}$ east
of north (Figure 8b).  
We also
require the inclination
to be fixed at 30$^{\circ}$.
This inclination
may be reasonable for the disk of
NGC 7714 itself \markcite{paperI}(Paper I),
but may well differ at larger radii as indicated by
the observed distortions to
the velocity field.
Therefore,
in Figure 9 we only plot the rotation
curve out to a radius of 
30$''$.
This curve shows a smooth increase in velocity
to a radius of 16$''$ where it drops off slightly.

\subsection{HI Absorption }

In Figure 10, we plot the continuum-subtracted
HI spectrum for the central position 
of NGC 7714.  
A weak absorption line
with an optical depth of 0.062 $\pm$ 0.016 and a velocity
width of 40 km s$^{-1}$
is apparent at a velocity
redshifted 100 km s$^{-1}$ with respect to the systemic
velocity of the galaxy.
This redshift suggests that gas is infalling onto
the starburst nucleus.
Assuming a spin temperature of 100K, 
we find an absorbing column density N(HI)
of 4.7 $\pm$ 1.2 $\times$ 10$^{20}$ cm$^{-2}$ towards the nucleus. 
Redshifted HI absorption lines
towards nuclear radio continuum sources are not uncommon,
and are more often redshifted than blueshifted 
relative to the systemic velocity of the galaxy
\markcite{d86}(Dickey 1986;
\markcite{v89}van Gorkom et al. 1989;
\markcite{s94}Smith 1994).

No HI absorption is seen towards the background continuum sources.

\section{An Off-Center Collision between Two Disk Galaxies}

Both the stars and the gas in this system 
appear to be distributed in a highly nonequilibrium and
short-lived manner.  Moreover, there are large
differences in the distribution of the two components.
To understand the dynamical processes that
produced these morphologies and to relate this
system to other collisional galaxies, detailed
comparisons between the data and numerical simulations
of galaxy interactions are important.
In \markcite{PaperI}Paper~I, we presented  
a restricted 3-body model of this interaction
which reproduced the stellar structure of
this system.
To understand the behavior of the gas, however,
a more sophisticated numerical model which includes
hydrodynamical effects is required.
Unfortunately, encounters with
interaction parameters
appropriate for this system
have not yet been investigated with hydrodynamical modeling
studies. 
Previously published hydrodynamical models of interacting
galaxies have concentrated on either
small impact parameter collisions 
(r$_{min}$ $<$ 0.5 r$_{disk}$) 
\markcite{as87}(Appleton 
$\&$ Struck-Marcell 1987; 
\markcite{sa87}Struck-Marcell $\&$ Appleton 1987;
\markcite{glb92}Gerber, Lamb, $\&$ Balsara 1992, 
\markcite{glb96}1996;
\markcite{sh93}Struck-Marcell $\&$ Higdon 1993;
\markcite{hw93}Hernquist $\&$ Weil 1993;
\markcite{acs96}Appleton et al. 1996; 
\markcite{s96a}Struck 1996a, \markcite{s96b}1996b) or
encounters without an initial collision 
(r$_{min}$ $\ge$ 1 r$_{disk}$) 
(\markcite{bh92}Barnes
$\&$ Hernquist 1992; 
\markcite{h93}Howard et al. 1993;
\markcite{mh96}Mihos $\&$ Hernquist 1996), while
our previous study of NGC 7714/7715 indicates that
an impact parameter between these two ranges is needed
\markcite{PaperI}(Paper~I).

To investigate the behavior of interstellar gas in such
an encounter,
in this section we present a numerical
simulation of 
an off-center collision
between two disk galaxies using
a smoothed particle hydrodynamics (SPH) code.
This model is not an
exact representation of NGC 7714/7715 itself, but
rather a more general off-center collision
which illustrates some of the effects
observed in NGC 7714/7715.  
We are currently working to produce a more
detailed model matched to NGC 7714/7715.

The SPH
code used is described in detail by 
\markcite{s96a}Struck (1996a, \markcite{s96b}1996b).
Earlier versions of this code
were used to model
the Cartwheel \markcite{sh93}(Struck-Marcell
$\&$ Higdon 1993) and VII Zw 466 
\markcite{acs96}(Appleton et al. 1996) systems.
The large-scale collisional dynamics are simulated with a
restricted 3-body approximation, while local self-gravity
between neighboring gas particles is included.
Simplified treatments of gas heating and cooling
are also included.  
Stars are represented as collisionless gas particles
which do not participate in heating or cooling.
The gas/star particle ratio is about 3/1 and the stellar disk
is half the size of the gaseous disk.
The primary galaxy contains 13,600 gas particles,
while the companion has 4000.

In this simulation, we used a galaxy mass ratio
M$_2$/M$_1$ = 0.3 and an impact parameter 
r$_{min}$ = 0.98 r$_{stellar~disk}$ = 0.5 r$_{gas~disk}$.
The inclination of the companion's orbit relative to the
main 
disk = 120$^{\circ}$
and the inclination
of the larger disk
relative to the plane of the sky = 30$^{\circ}$.
The initial orientation of the companion
is 25$^{\circ}$ out
of the x$-$z plane (the x$-$y plane is assumed to
be the plane of the sky).
As in the Paper I model,
the encounter is retrograde with 
respect to the main galaxy and prograde with
respect to the smaller galaxy.
In contrast to the models in 
\markcite{s96a}Struck (1996a, \markcite{s96b}1996b),
for this simulation
we used a softened point mass potential rather than 
a rigid halo.

Our final model is shown in Figure 11.  The four
panels show different particle sets, all at the same timestep.  
Figure 11a shows the stellar distribution in the x$-$y plane, while
Figures 11b and 11c show 
two orthogonal views of the
gas distribution, in the x$-$y and x$-$z planes.
In these figures,
the near side of the larger disk
is to the left and its rotation is clockwise on the
sky,
while
the lower side of the smaller galaxy 
is closer, with counterclockwise rotation.
Figure 11d shows an expanded x-y view of the bridge,
with both stars and gas.
The time displayed is 10$^8$ yrs after the point of closest approach,
assuming a mass of about 1.2 $\times$ 
10$^{10}$ M$_{\sun}$ contained within
the main stellar disk.

Following impact, the stellar morphology of this model developed in
a manner very similar to that in the 
\markcite{PaperI}Paper I model.
Specifically, a ring and two projecting
spiral arms develop in the primary stellar disk,
while stars from the companion form a stellar bridge
and a countertail.
The gas distribution is qualitatively similar to
that of the stars, but with some
important differences. 
The gas arms are longer, as expected for
the more extended gas disk.  
There is also a partial ring of gas in
the disk of the main galaxy, which lies somewhat to the outside
of the stellar ring.
Gas is also found in the bridge between the
two galaxies, but, unlike 
the stars in the bridge, this gas originates in both galaxies.
As in the 
\markcite{s96a}Struck (1996a, \markcite{s96b}1996b)
models,
gas has been splashed from both galaxies into the bridge
by the collision of the two disks.

By varying the initial orbital, disk, and viewing parameters,
we are able to change the position of the stellar bridge
relative to the gaseous bridge.  In particular, the 
location of the stellar bridge is determined in part
by the initial orientation of the companion galaxy,
while the impact position and rotation of the main
disk affect the location of the splash bridge.
In the model shown,
an offset between the
centroids of the gas bridge and the stellar bridge is apparent,
with the stars shifted to the top of the page
relative to the gas.
This offset is caused by dissipative collisions
between gas clouds
during the encounter; the gas from the companion
becomes displaced from
the stars.

\section{Discussion}

\subsection{The Gas/Star Offset in the Bridge}

One of the most striking results of
this study is the observation of a
very gas-rich bridge connecting NGC 7714 and NGC 7715.
In this bridge, the gas and stars are offset by
$\sim$2 kpc.
The HI column density in this bridge ($\sim$5 $\times$
10$^{21}$ cm$^{-2}$) is very high compared
to other such features.
It is about an order of magnitude higher than 
that observed in the Cartwheel 
\markcite{h96}(Higdon 1996)
and VII Zw 466 
\markcite{acs96}(Appleton et al. 1996) bridges
and in many tidal tails (e.g., 
\markcite{h94}Hibbard et al.
1994; 
\markcite{s94}Smith 1994; 
\markcite{hv96}Hibbard $\&$ van Gorkom 1996),
but is comparable to the HI column density observed
in the bridge of the `Taffy' galaxies, UGC 12914/5,
another possible bridge/ring system 
\markcite{c93}(Condon et al. 1993).

The NGC 7714/7715 bridge is clearly not a purely
`splash' bridge, as in the models of 
\markcite{s96a}Struck
(1996a, \markcite{s96b}1996b), because of the existence
of the pronounced stellar bridge.
It is also probably not a purely tidal bridge;
the presence of a stellar ring in NGC 7714 suggests 
an impact parameter $<$ r$_{disk}$ and therefore
a collision of the two gaseous disks.
Our hydrodynamical model of an off-center collision
between two disk galaxies (Section 4) produces
a connecting bridge with both stars and gas.
The bridge gas in this model originated in both
galaxies, while the stars came from only the
companion.  
We therefore suggest that the bridge in NGC 7714/7715
is a combination of a tidal bridge and
a `splash' bridge, where the stellar bridge has
been drawn out by tidal forces and 
both tidal and hydrodynamical processes created
the gaseous bridge.
The observed separation between the gas and stars
in the NGC 7714/7715 bridge is produced naturally
in such a scenario, because the stars and gas in
the bridge did not all originate from the same galaxy
and because the cross-section of gas is much larger
than the cross-section of stars.

Other alternative explanations for the gas/star
shift in the NGC 7714/7715 bridge are less likely.
An interaction with an intergalactic medium may
cause an offset,
however, NGC 7714/7715 is 
an isolated
pair, not in a cluster, so one would not
expect significant stripping by intergalactic gas.
Gas depletion
by star formation on the southern edge
of bridge could also create the offset.
It is likely, however, that the stars in
the stellar bridge predate the interaction
and were not formed in situ.
This possibility should be tested with
broadband optical and near-infrared mapping of
the bridge.

Gas/star offsets are not uncommon in tidal features.
A gas/star shift is observed
along about one quarter of the 3$'$ 
(80 kpc) 
long 
Arp 295 bridge
\markcite{hv96}(Hibbard $\&$ van Gorkom 1996),
and similar offsets between the stars and gas 
have been seen in
the tidal tails of a few interacting
systems
\markcite{w84}(Wever et al. 1984; \markcite{hv96}Hibbard
$\&$ van Gorkom 1996).
Although these are not ring galaxies, 
a similar mechanism may be responsible for the observed offsets 
in some cases: collisions may occur between clouds from
the two galaxies when the approach distance is small enough.
In one of the larger impact parameter models of 
\markcite{mh96}Mihos $\&$ Hernquist (1996),
a pronounced 
offset between the stars and gas in the bridge
is produced by collisions between clouds on intersecting
orbits.
In many tails no offset is seen
\markcite{hv96}(Hibbard 
$\&$ van Gorkom 1996), so observing such an offset may
depend on the
geometry of the encounter, the gas content,
and the timescale.
The NGC 7714/7715 interaction is very favorable
for the observation of this effect, because
of the relatively small impact parameter 
(r$_{min}$ $<$ r$_{disk}$)
compared to non-colliding interacting pairs.

\subsection{Gas Accretion From the Bridge}

In both the tidal 
\markcite{hb91}(Hernquist $\&$ Barnes 1991;
\markcite{mh96}Mihos $\&$ Hernquist 1996)
and collisional 
\markcite{s96a}(Struck
1996a, \markcite{s96b}1996b)
models of bridge formation,
accretion from the bridge onto the 
galaxies is expected.
In our HI data, we see deviations from circular
motion at the bridge, but, as noted in Section 3.4,
because of the uncertain geometry of the system
it is unclear whether these deviations
imply radial motion or simply non-planar
circular motion.
The interaction is mainly occuring in the
plane of the sky, and it is uncertain whether
NGC 7714 or NGC 7715 is closer to us.
Therefore, 
our data are inconclusive on the question
of gas accretion onto the galaxies from the bridge.

\subsection{The Rings }

NGC 7714 has two partial rings:
a stellar ring to the east, and 
a larger gaseous ring to the northwest.
As noted previously, these rings differ
from those in many ring galaxies in not 
having on-going star formation.
Like the non-star-forming inner ring
of the Cartwheel but unlike 
the Cartwheel's outer ring \markcite{h96}(Higdon 1996),
the optical ring in NGC 7714 does
not have a prominent HI 
counterpart.
Instead, HI depressions are observed on
the outer edge of the ring and at the southern
end of the bar.  
Our hydrodynamical model of an off-center collision
(Section 4)
shows that such differentiation of gas and stars
is expected in partial rings formed during
very off-center collisions.
Thus the lack of an HI counterpart 
does not rule out a collisional
origin for the NGC 7714 ring.
This point will be investigated further in future
modeling studies.

One of the best methods of testing the collisional
ring hypothesis 
is to search for 
kinks in the velocity field across the rings.
In the 
more symmetrical ring galaxy the Cartwheel,
such kinks are observed and are consistent with
expansion 
\markcite{h96}(Higdon 1996).  
In NGC 7714, the data are inconclusive on this point.
The isovelocity contours do curve slightly when crossing
the stellar ring (Figure 8a), and the gas is slightly blueshifted
at the ring relative to circular motion, implying
expansion, if one assumes the NGC 7714 tails
are trailing.  
However, interpretation is complicated
by the fact that large peculiar motions are seen in
the nearby bridge, making it hard to separate possible
radial or non-planar motions in the bridge from those in the ring.
Also, the fact that the gas in the vicinity of the stellar ring is
not confined to a ring-like structure makes interpretation
of its motions uncertain.  
The isovelocity curves across 
the northwestern loop
also
do not
show pronounced kinks, thus neither ring
shows clear evidence for expansion.  
The fact that these are 
asymmetric rings likely caused by a very off-center encounter
may also
contribute to the lack of clearly defined velocity deviations
at the rings.

Another possibility is that the stellar ring in this system
is simply a wrapped-around spiral arm caused by
a non-collisional prograde planar encounter, as in
some of the models of 
\markcite{tw90}Thomson
$\&$ Wright (1990), 
\markcite{des91}Donner, 
Engstr\"om, $\&$ Sundelius
(1991), 
\markcite{e91}Elmegreen et al. (1991), and
\markcite{h93}Howard et al. (1993), rather than a collisional ring.
The high column density of gas in the bridge, however, and
its offset from the stars,
supports the idea that
gas was forced out of the main galaxy by a collision
between two gas disks rather than merely perturbed in
a grazing or long-range encounter.   This possibility should be
investigated further with hydrodynamical simulations of
such encounters.

\section{Conclusions}

In this paper, we have presented both optical
and 21 cm HI maps of the interacting pair NGC 7714/7715.
This pair is connected by two bridges: a gaseous bridge,
in which star formation is on-going, and a stellar
bridge, which is parallel to but displaced 10$''$ (1.8 kpc)
to the south of 
the gaseous bridge.
This offset persists 2\farcm7 (30 kpc) from 
the bridge through NGC 7715 and the countertail east of NGC 7715.
In the main body of NGC 7714, there is no prominent
HI counterpart to the stellar ring.  Instead, we 
find a clumpy interstellar medium with maxima near 
the prominent star formation complexes in the disk, and
a prominent HI loop on the opposite side of the galaxy.
We also detect
redshifted
HI absorption towards the 
NGC 7714 nucleus, suggesting
infall onto the starburst nucleus.  

The velocity field of the inner disk of NGC 7714 shows
the standard `spider diagram' signature of an
inclined rotating disk.  At radii larger than 20$''$ (3.5 kpc)
pronounced deviations from circular motion are present,
particularly at the bridge, indicating nonplanar and/or
radial motions.  Signs of rotation are also observed at
NGC 7715.  The gas in the immediate vicinity
of NGC 7715 appears kinematically
distinct from the gas in the surrounding extended
structures.

We suggest that the peculiar morphology of this
system is the result of a very off-center collision
between two gas-rich 
disk galaxies.
Our numerical simulations indicate that such
collisions can create bridge/ring systems similar
to NGC 7714/7715, in which the bridge contains
both gas and stars.
We conclude that the bridge in NGC 7714/7715 is a hybrid between
classical tidal bridges, such as that in seen in M51,
and gaseous `splash' bridges observed in ring galaxies
such as the Cartwheel.
The observed offset between
the gas and stars in this bridge
may have been caused by dissipative collisions
between gas clouds from the two galaxies.
Further investigations of 
the interaction parameters for the NGC 7714/7715 encounter
are needed to better define the 
roles of tidal `swing' and collisional
`splash' in the NGC 7714/7715 bridge, and to better match
the details of the system.

\acknowledgments

We would like to thank the staff at NRAO for their
help in obtaining and reducing the data.  In particular, 
James Higdon provided much assistance with the data reduction and
many helpful comments.
Data from the Digitized Sky Survey were used in this study.
This research has made 
use of the NASA/IPAC Extragalactic Database
(NED) which is operated by the Jet Propulsion Laboratory, California 
Institute of Technology, under contract with the National
Aeronautics and Space Administration.
This research was supported in part by NASA grant NAG 2-67.
A portion of this work was performed while B.J.S.
held a National Research Council Research Associateship
at the Jet Propulsion Laboratory.
R.W.P. was supported in part by NSF grant AST-9112879.

\clearpage

{\bf Captions}

Figure 1. The B+C+D Array channel maps for NGC 7714/7715.
The beamsize is 11\farcs02 $\times$ 8\farcs48 with a position
angle of $-$36\fdeg6 and the noise level is 0.3 mJy beam$^{-1}$.
The first contour is 1.28 mJy beam$^{-1}$; the contour
interval is 2.56 mJy beam$^{-1}$.
The crosses indicate the locations of the nuclei of the two galaxies.

Figure 2. (a) 
The narrowband red continuum ($\lambda$ = 6560$\AA$;
$\Delta$$\lambda$ = 50$\AA$) image of the NGC 7714/7715 system.
NGC 7715 lies to the east of NGC 7714.
The contours are logarithmically spaced, with
$\Delta$log(F) = 0.12.
These data have been smoothed to 3$''$ resolution for clarity.
The brightening in the lower left corner of
this figure is due to a very bright star near the edge
of the image.
(b) The H$\alpha$ image of the NGC 7714/7715 system.
NGC 7714 and the HII regions in the bridge are visible
in this image.  NGC 7715 
is not seen in this image
because it does not
have on-going star formation.
The first contour is 3.9 $\times$ 10$^{-16}$ erg s$^{-1}$ cm$^{-2}$ (3$\sigma$).
The contour interval is logarithmic,
with $\Delta$log(F) = 0.5.
The features outside of NGC 7714 and the bridge are
artifacts caused by imperfect stellar subtraction.

Figure 3. The B Array 20 cm continuum map. The beamsize
is 6\farcs32 $\times$ 5\farcs79 with a positions angle
of 2\fdeg7, while the rms noise
level is 0.18 mJy/beam.
The first contour level is 0.72 mJy/beam.
The contour interval is logarithmic, with $\Delta$log(F) = 0.5.

Figure 4. (a) The naturally-weighted B Array HI intensity map.
The spatial resolution is 6\farcs32 $\times$ 5\farcs78.
The first contour level and the contour interval
are 4 $\times$ 10$^{20}$ cm$^{-2}$.
(b) The naturally-weighted B+C+D Array HI intensity
map.
The beamsize is 11\farcs02 $\times$ 8\farcs48 with
a position angle of $-$36\fdeg6.
The first contour level is 
2 $\times$ 10$^{20}$ cm$^{-2}$
and the contour interval
is 4 $\times$ 10$^{20}$ cm$^{-2}$.

Figure 5. (a) The B Array HI intensity map (greyscale)
(6\farcs32 $\times$
5\farcs78 beam), superposed on the narrowband red continuum
image (contours).
(b) The B Array HI intensity map (greyscale)
(6\farcs32 $\times$
5\farcs78 beam), superposed 
on the H$\alpha$ contours.

Figure 6. The POSS-I red image, smoothed to 12$''$
resolution (contours),
superposed on the B+C+D Array HI intensity map (greyscale).
A bright star in the lower left corner of the optical
image causes a background gradient across the image.

Figure 7. HI spectra for the various structures
in the NGC 7714/7715 system, as in Table 3.
Note that the NGC 7715 spectrum has been scaled up by a factor
of four.

Figure 8. (a) The B+C+D Array
velocity field
(contours), superposed on the red continuum image
(greyscale).  The contour interval is 10 km s$^{-1}$.
Several contours are labeled.
(b) The velocity field
(contours), superposed on the B+C+D
Array HI intensity map
(greyscale).  The contour interval is 10 km s$^{-1}$.
Several contours are labeled.

Figure 9.  The rotation curve for NGC 7714.  This
was derived assuming a constant inclination of 30$^{\circ}$,
a systemic velocity of 2800 km s$^{-1}$,
and a line of nodes of 136$^{\circ}$ east of north.

Figure 10.  The continuum-subtracted HI spectrum
for the nucleus of NGC 7714,
from the B Array data. 

Figure 11.  A numerical simulation 
of an off-center collision between two unequal mass disk
galaxies (see text),
at a time 10$^8$ yrs after the point of
closest approach.  
This model was run with a smooth particle hydrodynamics
code as described in the text.
Panel a shows an x$-$y view of the distribution of stars.
In panels b and c, the distribution of gas in the
x$-$y and x$-$z planes, respectively, are displayed.
Panel d gives an expanded x$-$y view of both the gas
and stars in the bridge.
In panels a-c, every 10th gas particle and every 3rd star particle are 
plotted.  In panel d, every 5th gas particle and
every 2nd star are plotted, and the stars are 
plotted as plus signs.

\clearpage

\begin{center}
{\bf Table 1}\\
Basic Information on the NGC 7714/7715 System\\ [12pt]
\begin{tabular}{cccccccclclcr} \hline
Galaxy&NGC 7714&NGC 7715&Notes\\
$\alpha$(1950)&
23$^{\rm h}$33$^{\rm m}$$40\rlap.^{\rm s}6$&
23$^{\rm h}$33$^{\rm m}$$48\rlap.^{\rm s}3$&$^a$\\
$\delta$(1950)&
1\arcdeg 52\arcmin 42\arcsec
& 1\arcdeg 52\arcmin 48\arcsec&$^a$\\
Type&SBb Pec&Irr Pec&$^b$\\
B$_{T}$&13.0&14.7&$^b$\\
Velocity&2808 $\pm$ 15 km s$^{-1}$&2812 $\pm$ 31 km s$^{-1}$&$^c$\\ \hline
\end{tabular}
\end{center}
$^a$For NGC 7714, radio continuum
position from Condon et al. 1990.
For NGC 7715, optical position measured with the Grant machine at N.O.A.O. (Smith 1989).\\
$^b$de Vaucouleurs et al. 1991.\\
$^c$Optical velocities tabulated in de Vaucouleurs et al. 1991.\\

\clearpage

\begin{center}
   {\bf Table 2}\\
   Background 20 cm Continuum Sources in the NGC 7714 field\\ [12pt]
   \begin{tabular}{lllllllcccclccclclclcrcccc} \hline
       \multicolumn{1}{c}{Source}&
\multicolumn{3}{c}{R.A.}
&\multicolumn{3}{c}{Dec.}
 &\multicolumn{1}{c}{Peak $F_{20}$$^a$}\\
      \multicolumn{1}{c}{}&\multicolumn{3}{c}{(1950)}&
      \multicolumn{3}{c}{(1950)}&
      \multicolumn{1}{c}{(mJy)}& \\ \hline
$\#$1&23$^{\rm h}$&33$^{\rm m}$&57\fs54&
1$^{\circ}$&54$'$&16$''$&30.5\\
$\#$2&23$^{\rm h}$&33$^{\rm m}$&57\fs14&
1$^{\circ}$&54$'$&~8$''$&42.7\\
$\#$3&23$^{\rm h}$&33$^{\rm m}$&56\fs34&
1$^{\circ}$&53$'$&56$''$&54.8\\
\\
   \end{tabular}
\end{center}
$^a$In the 6\farcs32 $\times$ 5\farcs79 naturally-weighted
B Array beam.

\clearpage

\begin{center}
   {\bf Table 3}\\
   HI Structures in the NGC 7714/7715 System$^a$\\ [12pt]
   \begin{tabular}{lccccccccccccc} \hline
\multicolumn{1}{c}{Source}&
\multicolumn{3}{c}{R.A.$^b$}
&\multicolumn{3}{c}{Dec.$^b$}
 &\multicolumn{1}{c}{Velocity$^c$}
 &\multicolumn{1}{c}{$\Delta$V$^d$}
 &\multicolumn{1}{c}{M$_{HI}$}
 &\multicolumn{1}{c}{peak N(HI)$^e$}\\
      \multicolumn{1}{c}{}&
      \multicolumn{3}{c}{(1950)}&
      \multicolumn{3}{c}{(1950)}&
\multicolumn{1}{c}{(km s$^{-1}$)}&
\multicolumn{1}{c}{(km s$^{-1}$)}&
\multicolumn{1}{c}{(M$_{\sun}$)}& 
\multicolumn{1}{c}{(cm$^{-2}$)}& \\ \hline
Eastern Cloud&23&33&51.1&1&53&12&2750&40&
3.8 $\times$ 10$^8$&1.8 $\times$ 10$^{21}$\\
Bridge&23&33&44.9&1&52&50&2790&50&
1.5 $\times$ 10$^9$&4.2 $\times$ 10$^{21}$\\
Northwestern Loop&23&33&42.7&1&53&32&2880&120&
1.1 $\times$ 10$^9$&1.7 $\times$ 10$^{21}$\\
Southwestern Clump&23&33&39.5&1&52&28&2820&70&
2.9 $\times$ 10$^8$&4.1 $\times$ 10$^{21}$\\
Northwestern Clump&23&33&40.7&1&52&52&2850&80&
2.4 $\times$ 10$^{8}$
&3.8 $\times$ 10$^{21}$
\\
Southeastern Arc&23&33&40.9&1&52&24&2760&60&
2.1 $\times$ 10$^8$&2.8 $\times$ 10$^{21}$\\
Total System&23&33&44.9&1&52&50&2800&140&
7.0 $\times$ 10$^9$&4.2 $\times$ 10$^{21}$\\
NGC 7714$^f$&23&33&39.5&1&52&28&2830&150&1.7 $\times$ 10$^9$&
4.1 $\times$ 10$^{21}$\\
NGC 7715$^g$&23&33&49.1&1&52&58&2760&80&3.1 $\times$
10$^8$&1.9 $\times$ 10$^{21}$\\
\\
   \end{tabular}

\end{center}
$^a$All results from the B+C+D Array data.\\
$^b$Position of peak HI column density.\\
$^c$Peak velocity.\\
$^d$Full Width Half Maximum.\\
$^e$In a 11\farcs02 $\times$ 8\farcs48 beam.\\
$^f$For a 50$''$ $\times$ 50$''$ region centered
on the NGC 7714 nucleus.  Includes the two clumps
and the southeastern arc listed above.\\
$^g$For a 30$''$ $\times$ 30$''$ region centered
on the NGC 7715 nucleus.\\


\clearpage


\begin{references}

\reference{as87} Appleton, P. N., $\&$ Struck-Marcell, C. 1987,  \apj,
318, 103

\reference{acs96} Appleton, P. N., Charmandaris, V., $\&$ Struck, C.
1996, \apj, 468, 532.

\reference{a66} Arp, H. C. 1966, Atlas of Peculiar Galaxies, (Pasadena: California
Institute of Technology)

\reference{bh91} Barnes, J. E., $\&$ Hernquist, L. 1991, \apj, 
370, L65

\reference{b89} Begeman, K. G. 1989, \aap, 223, 47

\reference{b93} Bernl\"ohr, K. 1993, \astap, 268, 25

\reference{b94} Bowen, D. V., Osmer, S. J., Blades, J. C., Tytler, D., Cottrell, L.,
Fan, X.-M., $\&$ Lanzetta, K. M. 1994, \aj, 107, 461

\reference{b87} Bushouse, H. A. 1987, \apj, 320, 49

\reference{bw90} Bushouse, H. A., $\&$ Werner, M. W. 1990, \apj, 359,
72

\reference{c80} Clark, B. G. 1980, \astap, 89, 377

\reference{c90} Condon, J. J., Helou, G., Sanders, D. B., $\&$ Soifer, B. T.
1990, \apjsupp, 73, 359

\reference{c93} Condon, J. J., Helou, G., Sanders, D. B.,
$\&$ Soifer, B. T. 1993, \aj, 105, 1730

\reference{ds88} De Robertis, M. M., $\&$ Shaw, R. A. 1988, \apj, 329,
629

\reference{RC3} de Vaucouleurs, G., de Vaucouleurs, A., Corwin, H. G., Jr.,
Buta, R. J., Paturel, G., $\&$ Fouque, P. 1991, Third Reference
Catalogue of Bright Galaxies, (New York: Springer-Verlag) (RC3)

\reference{d86} Dickey, J. M. 1986, \apj, 300, 190

\reference{des91} Donner, K. J., Engstr\"om, S., $\&$
Sundelius, B. 1991, \aap, 252, 571

\reference{e91} Elmegreen, D. M., Sundin, M., Elmegreen,
B., $\&$ Sundelius, B. 1991, \aap, 244, 52

\reference{fh77} Fosbury, R. A. E., $\&$ Hawarden, T. G. 1977, \mnras,
178, 473

\reference{f80} French, H. B. 1980, \apj, 240, 41

\reference{glb92} Gerber, R. A., Lamb, S. A., $\&$ Balsara, D. S.
1992, \apj, 399, L55

\reference{glb96} Gerber, R. A., Lamb, S. A., $\&$ Balsara, D. S.
1996, \mnras, 278, 345

\reference{g95} Gonz\'alez-Delgado, R. M., P\'erez, E.,
D\'ias, \'A. I., Garci\'a-Vargas, M. L., Terlevich, E.,
$\&$ Vilchez, J. M. 1995, \apj, 439

\reference{hw93} Hernquist, L., $\&$ Weil, M. L. 1993, \mnras, 261, 804

\reference{h93} Hibbard, J. E., Guhathakurta, P., van
Gorkom, J. H., $\&$ Schweizer, F. 1994, \aj, 107, 67

\reference{hv96} Hibbard, J. E., $\&$ van Gorkom, J. H. 1996,
\aj, 111, 655.

\reference{h95} Higdon, J. L. 1995, \apj, 455, 524

\reference{h96} Higdon, J. L. 1996, \apj, 467, 241

\reference{h93} Howard, S., Keel, W. C.,
Byrd, G., $\&$ 
Burkey, J. 1993, \apj, 417, 502

\reference{hms56} Humason, M. L., Mayall, N. U., $\&$ Sandage, A. R.
1956, \aj, 61, 97

\reference{i93} Ishizuki, S. 1993, Ph.D. Thesis,
University of Tokyo

\reference{k84} Keel, W. C. 1984, \apj, 282, 75

\reference{lt76} Lynds, R., $\&$ Toomre, A. 1976, \apj, 209, 382

\reference{mb88} Maloney, P., $\&$ Black, J. H. 1988, \apj, 325, 389

\reference{mah92} Marcum, P. M., Appleton, P. N., $\&$ Higdon, J. L.
1992, \apj, 399, 57

\reference{ma95} Marston, A. P., $\&$ Appleton, P. N. 1995, AJ,
109, 1002

\reference{mh96} Mihos, J. C., $\&$ Herquist, L. 1996, \apj, 464, 641
 
\reference{ms88} Mirabel, I. F., $\&$ Sanders, D. B. 1988, \apj, 335, 104

\reference{nhg89} Neff, S. G., Hutchings, J. B., $\&$ Gower, A. C. 1989, \aj, 97, 1291

\reference{ps74} Peterson, S. D., $\&$ Shostak, G. S. 1974, \aj,
 79, 767

\reference{sss91} Sanders, D. B., Scoville, N. Z., $\&$ Soifer, B. T. 1991, \apj,
370, 158

\reference{s78} Schweizer, F. 1978, in IAU Symposium 77, Structure
and Properties of Nearby Galaxies, ed. E. M. Berkhuijsen and R. Wielebinski
(Dordrecht: Reidel), p. 279

\reference{paperI} Smith, B. J., $\&$ Wallin, J. F. 1992, \apj, 393, 544

\reference{s94} Smith, B. J. 1994, \aj, 107, 1695

\reference{s89} Smith, B. J. 1989, Ph.D. Disssertation, University
of Massachusetts

\reference{s87} Soifer, B. T., Sanders, D. B., Madore, B. F., Neugebauer,
G., Danielson, G. E., Elias, J. H., Lonsdale, C. J., $\&$
Rice, W. L. 1987, \apj, 320, 238

\reference{s90} Struck-Marcell, C. 1990, \aj, 99, 71

\reference{sa87} Struck-Marcell, C., $\&$ Appleton, P. N. 1987, \apj,
323, 480

\reference{sh93} Struck-Marcell, C., $\&$ Higdon, J. L. 1993, \apj, 411, 108

\reference{s96a} Struck, C. 1996a, \apj, submitted

\reference{s96b} Struck, C. 1996b, \apj, submitted

\reference{s96c} Struck, C., Appleton, P. N.,
Borne, K. D., $\&$ Lucas, R. A. 1996, \aj, in press

\reference{s93} Surace, J. A., Mazzarella, J., Soifer, B. T., $\&$ Wehrle, A. E.
1993, \aj, 105, 864

\reference{ts77} Theys, J. C., $\&$ Spiegel, E. A. 1977, \apj, 212, 616

\reference{th90} Thomson, R. C., $\&$ Wright, A. E.
1990, \mnras, 247, 122

\reference{tt72} Toomre, A., $\&$ Toomre, J. 1972, \apj, 178, 623

\reference{v85} van Breugel, W., Filippenko, A. V., Heckman, T. M., 
$\&$ Miley, G. K.
1985, \apj, 293, 83

\reference{v89} van Gorkom, J. H., Knapp, G. R., Ekers, R. D.,
Ekers, D. D., Laing, R. A., $\&$ Polk, K. S.
1989, \aj, 97, 708

\reference{ws94} Wallin, J. F., $\&$ Struck-Marcell, C. 1994, \apj,
433, 631

\reference{w81} Weedman, D. W., 
Feldman, F. R., Balzano, V. A., Ramsey, L. W., Sramek,
R. A., $\&$ Wu, C.-C. 1981, \apj, 248, 105

\reference{w84} Wever, B. M. H. R., Appleton, P. N.,
Davies, R. D., $\&$ Hart, L. 1984, \aap, 140, 125

\reference{w72} Wright, A. E. 1972, \mnras, 157, 309

\reference{yd91} Young, J. S., $\&$ Devereux, N. A. 1991, \apj, 373, 414

\end{references}
\end{document}